\documentclass[a4paper]{jpconf}
\usepackage{graphicx}
\begin{document}

\title{Jet quenching in relativistic heavy ion collisions}

\author{Ivan Vitev}

\address{Los Alamos National Laboratory, Theory Division 
and Physics Division, 
Mail Stop H846 \\ Los Alamos, NM 87545, USA}

\ead{ivitev@lanl.gov}

\begin{abstract}

Parton propagation in dense nuclear matter results in elastic, 
inelastic and coherent multiple soft scattering with the in-medium 
color charges. Such scattering leads to calculable modifications of 
the hadron production cross section that is evaluated in the framework 
of the perturbative QCD factorization approach. Final state 
medium-induced gluon bremsstrahlung is arguably the most 
efficient way of suppressing large transverse momentum 
particle production in nucleus-nucleus collisions. The observed 
hadronic attenuation, known as jet quenching, can be related to 
the properties of the medium, such as density and temperature, and 
carries valuable information about the early stages of heavy ion 
reactions. Non-Abelian energy loss in the quark-gluon plasma can be 
studied in much greater detail through the modification of the two 
particle back-to-back correlations. Perturbative calculations give 
good description of the redistribution of the lost energy in lower 
transverse momentum particles and predict significant increase of 
the correlation width of away-side di-hadrons. In contrast, energy 
loss in cold nuclear matter was found to be small but for large 
values of Feynman-x is expected to complement the dynamical higher twist 
shadowing in experimentally observable forward rapidity hadron 
suppression.

\end{abstract}

\section{Energy loss in hot and dense nuclear matter}

Jet quenching, the suppression  of high transverse 
momentum hadron production relative to the expectation from
p+p collisions scaled by the number of elementary  nucleon-nucleon 
interactions, has been predicted~\cite{Wang:1991xy} 
to be one of the key signatures for the creation of a hot and dense
quark-gluon plasma (QGP) in ultra-relativistic nuclear collisions.  
Since,  the non-Abelian energy loss of fast quarks and 
gluons in dense nuclear matter has been studied in great detail 
in a variety of theoretical approaches~\cite{Gyulassy:2003mc}. 
Its key features can be illustrated on the example of the reaction
operator formalism~\cite{Gyulassy:2000er}, which systematically 
expands  energy loss in terms of the correlations between multiple
scattering centers.

\begin{figure}[!t]
\hspace*{0.8cm}
\includegraphics[width=2.6in,angle=0]{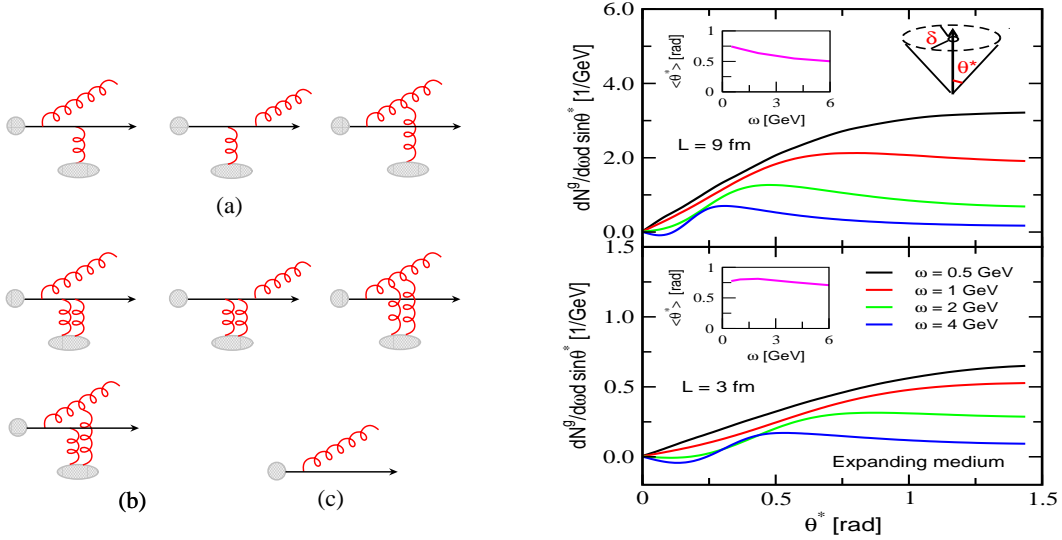} 
\hspace*{0.8cm}
\includegraphics[width=2.6in,height=2.8in,angle=0]{fig1.eps} 
\caption{ Left panel: diagrams contributing to the lowest 
order in opacity energy loss expansion.  Right panel: the angular  
distribution of medium-induced bremsstrahlung of $E = 6$~GeV gluon 
jet  for fixed values of the radiative gluon energy 
$\omega = 0.5, \, 1, \, 2, \, 4$~GeV. Top and bottom panels 
represent (1+1)D Bjorken expanding medium of transverse size 
$L=9$~fm and $L=3$~fm, respectively.
Inserts show  $\langle \theta^* \rangle$ versus $\omega$.}
\label{hole}
\end{figure}

For A+A collisions, the dominant final state energy loss follows
the hard partonic scattering, which takes place on a time scale
$t_0 \sim 1/Q$. Even in the absence of a medium, the large virtuality
leads to non-Abelian bremsstrahlung.  For real gluons of small 
and  moderate frequencies $\omega$~\cite{Gyulassy:2000er}  
\begin{equation} 
|{\cal{M}}_{c} |^2  \propto  
\frac{dN^g_{\rm vac}}{d\omega d \sin \theta^* d \delta } 
\approx  \frac{  C_R \alpha_s} { \pi^2} 
\frac{1}{ \omega \sin  \theta^*}  \;.
\label{divsmall} 
\end{equation}
In Eq.~(\ref{divsmall}) $\theta^* = \arcsin(k_\perp/\omega)$ is 
the angle relative to the jet axis,  $\delta$ is the azimuthal 
cone angle (both illustrated in right panel of Fig.~\ref{hole}),   
$\alpha_s$ is the strong coupling constant and 
$C_R = 3 \, (4/3)$ for gluon (quark) jets, respectively. 
Virtual gluon corrections remove the $\omega \rightarrow 0$ 
infrared singularity in the cross sections in accord with 
the  Kinoshita-Lee-Nauenberg 
theorem~\cite{Kinoshita:1962ur} but the collinear   
$\theta^* \rightarrow 0$  divergence has to be regulated or subtracted 
in the parton distribution functions (PDFs) and the fragmentation 
functions (FFs).

In contrast, the final state medium-induced bremsstrahlung spectrum 
is both collinear and infrared safe. To first order in the 
mean number of soft interactions in the plasma the distribution of 
gluons in angle and frequency 
can be written as~\cite{Vitev:2005yg}
\begin{eqnarray} 
\hspace*{-.5cm} 
&& \!\!\!\!\! 
|{\cal{M}}_{a} |^2 + 2 {\cal R}e {\cal{M}}_{b}^\dagger  {\cal{M}}_{c} 
\propto  \frac{dN^g_{\rm med}}{d\omega d \sin \theta^*  d \delta } 
\approx   
\frac{2 C_R \alpha_s}{  \pi^2 } 
\int_{z_0}^L \frac{d \Delta z}{\lambda_g(z)}  
\int_0^\infty  d q_\perp \, q_\perp^2 \frac{1}{\sigma_{el}}  
\frac{d \sigma_{el}}{d^2 q_\perp} (z)   \int_0^{2 \pi} d \alpha \;  
 \nonumber \\[.5ex]
&&  
\frac{ \cos \alpha }
{q_\perp^2 - 
2 q_\perp \omega \sin \theta^*  \cos \alpha + \omega^2 \sin^2 \theta^*} 
 \left[ 1 -  \cos \left( \frac{ (q_\perp^2 - 
 2 q_\perp \omega \sin \theta^* \cos \alpha +  \omega^2 \sin^2 \theta^*) 
\Delta z}{2 \omega} 
\right)  \right] \;. \qquad \nonumber \\ 
\label{unintspect} 
\end{eqnarray}
In Eq.~(\ref{unintspect}) 
$\alpha = \angle (\vec{k}_\perp,\vec{q}_\perp )$,  
$\lambda_g(z)$ is the position-dependent 
gluon  mean free path  and  $L$ is the transverse size of the medium. 
The momentum transfers $\vec{q}_\perp$ are distributed according 
to the normalized  elastic scattering cross section 
$ {\sigma_{el}^{-1}}  {d \sigma_{el}} / {d^2 q_\perp} (z) 
= {\mu^2_D(z)}/{\pi ( q_\perp^2 + \mu^2_D(z))^2} $.
In this model 
$\langle q_\perp^2 \rangle \propto \mu_D^2(z)$ and for a quark-gluon
plasma in local thermal equilibrium  $\mu_D^2(z) \sim 4 \pi \alpha_s T^2$. 
For  the case of (1+1)D dynamical Bjorken expansion of the QGP
$ \mu_D(z) = \mu_D(z_0) \left({z_0}/{z} \right)^{1/3} \!\!, 
\, \lambda_g(z) = \lambda_g(z_0) 
\left({z}/{z_0} \right)^{1/3}\!\!$. At small angles $\theta^*$ 
both the direct and double Born terms~\cite{Gyulassy:2000er} 
in Eq.~(\ref{unintspect})
are divergent. However, they come with different signs 
$ \propto (-i)^2 = -1, \,  \propto  i(-i) = 1$, equivalent 
to a phase difference $\pi$. The soft and 
collinear divergences  then exactly cancel~\cite{Vitev:2005yg}.

From Eq.~(\ref{unintspect})  the gluon distribution 
is not only finite when $\theta^* \rightarrow 0$  but vanishes 
on average  due to the uniform angular distribution of momentum 
transfers from the medium, 
$\int_0^{2\pi} d \alpha \, \cos \alpha = 0$. 
We have checked that for physical gluons of $k_\perp \leq \omega$ 
the cancellations discussed  here persist to all orders in the 
mean number of  scatterings~\cite{Gyulassy:2000er}.
The small frequency and small angle spectral behavior  
of $\, {dN^g_{\rm med}}/{d\omega d \sin \theta^*  d \delta } \,$ remains 
under perturbative control.  
Given the vastly different angular behavior of the vacuum and 
the medium-induced gluon bremsstrahlung, Eqs.~(\ref{divsmall}) 
and (\ref{unintspect}),  it is critical to identify
the phase space where cancellation of the color currents 
induced by the hard and soft scattering occurs. We fix the parameters
of the medium to
$\mu_D(z_0) = 1.5$~GeV and $\lambda_g(z_0) = 0.75$~fm
at initial formation time $z_0 = 0.25$~fm. Since small frequency 
emission is suppressed, we use only a moderate $\alpha_s = 0.25$. 
Triggering on high $p_{T_1}$ hadron  directs its parent parton ``c'' 
away from the medium and  places the collision point of the lowest order 
(LO) ${\rm ab} \rightarrow {\rm cd}$ underlying perturbative 
process~\cite{Owens:1986mp} 
close to the periphery of the nuclear overlap region.
Then, it is the back-scattered jet ``d'' that 
traverses the QGP.  For large nuclei, such as Au and Pb,  path lengths  
$L = 9 \; (3)$~fm are used to illustrate central (peripheral) 
collisions, respectively.  We limit gluon emission to the forward 
jet hemisphere,  $0 \leq \theta^* \leq \frac{\pi}{2}$.

The angular distribution of medium-induced radiation for 
$E = 6$~GeV gluon jet for select values of $\omega$ is shown in 
Fig.~\ref{hole}. We find that gluon emission is strongly suppressed 
within a cone of opening angle $\theta^* \simeq 0.25$~rad.   
The broad gluon distribution can be characterized by the mean 
emission angle
$$ \langle  \theta^*  \rangle = {\int_0^1 
\theta^* \frac{dN^g_{\rm med_{}}}{d\omega d \sin \theta^* } 
\,d\sin \theta^*  }      \left[ {\int_0^1
\frac{dN^g_{\rm med_{}}}{d\omega d \sin \theta^*_{} }_{} \, 
d\sin \theta^* }   \right]^{-1} \!\!\! \;,  $$
given in the insert of Fig.~\ref{hole}. We emphasize that the angular
spectrum is distinctly non-Gaussian and the  mechanism that determines
its form is
based on the destructive interference of color currents from hard and soft 
parton scattering rather than a slow random walk of the gluon in $\theta^*$.
The large angle distribution or radiative quanta is evident for both 
large and small frequencies~$\omega$.

In the case of heavy quark creation and propagation the   
vacuum~\cite{Dokshitzer:1995ev} and the Bertsch-Gunion 
radiation~\cite{Thomas:2004ie} are modified by the large
mass $M_q$.  The mass enters in the propagators as follows, 
$k_\perp^2 \rightarrow k_\perp^2  +  \omega^2 M_q^2/E^2$, and 
suppresses the gluon bremsstrahlung for  $\sin \theta^* \leq M_q/E$. 
Such depletion can only be effective if gluon radiation peaks
at $\theta^* \rightarrow 0$. Since for light quarks the 
final state medium-induced  non-Abelian bremsstrahlung 
is {\em not} dominated by small angle emission, see 
left panel of Fig.~\ref{hole},
effects other than  the ``dead cone'' control the energy 
loss pattern of heavy versus light quarks.

Qualitatively, for light quarks the behavior of the energy loss  
as a function  of the density and the size of the system can be 
summarized to first order in opacity as follows:
\begin{eqnarray}
 \Delta E  &\approx&  \int dz \, 
\frac{C_R\alpha_s}{2} \frac{\mu^2}{\lambda_{g}} \, z \, 
\ln \frac{2E}{\mu^2  L  }   
= \int dz \,  \frac{9 C_R \pi \alpha_s^3}{4} \rho^g(z) 
\, \ln \frac{2E}{\mu^2  L  }  \nonumber  \\[1ex]
&& \qquad \quad  \!\! = \left\{ \begin{array}{ll}  
\frac{9 C_R \pi \alpha_s^3}{8} \, \rho^g  L^2 
\, \ln \frac{2E}{\mu^2  L  } 
 \, ,  &  {\rm static} \\[1ex]
 \frac{9  C_R \pi \alpha_s^3}{4} 
\frac{1}{A_\perp}  \frac{dN^{g}}{dy}  L 
\,   \ln \frac{2E}{\mu^2 L }  
 \, , & (1+1)D   \end{array}  \right.   .
\label{deltae}
\end{eqnarray}
Eq.~(\ref{deltae}) assumes  energetic jets and neglects kinematic
bounds. Here, $A_\perp$ is the transverse size of nuclear matter.  
For static systems $ \Delta E $ depends 
quadratically on the nuclear size. For the case of longitudinal 
Bjorken expansion this dependence is reduced to 
linear~\cite{Gyulassy:2000er}
but  the energy loss is sensitive to the initial parton rapidity 
density $dN^g/dy$. Understanding the effective color charge 
density dependence of $\Delta E$  the is the key to 
jet tomography~\cite{Vitev:2002pf}.

The calculated energy loss can be incorporated in the perturbative
QCD factorization approach as a modification of the single and
double inclusive hadron production cross sections. To lowest order
in the double collinear limit these are given by:
\begin{eqnarray} 
\label{single}
\frac{ d\sigma^{h_1 }_{NN} }{ dy_1  d p_{T_1} }  
& = &  K \sum_{abcd}  p_{T_1}
\int \frac{dz_1}{z_1^2} D_{h_1/c}(z_1) 
\int d x_a   \frac{\phi_{a/N}(x_a) \phi_{b/N}(x_b)   }{x_a x_b \, S}  
\left[\frac{1}{x_a S +  {U}/{z_1} }\right]  \nonumber \\[.5ex] 
&& \times  2 \pi   \alpha_s^2   |\overline {M}_{ab\rightarrow cd}|^2  \, , 
\\
 \frac{ d \sigma^{h_1 h_2}_{NN} }{ dy_1  dy_2 
 dp_{T_1} dp_{T_2} d\Delta \varphi} & = & K
\sum_{abcd}  \int \frac{dz_1}{z_1} \, D_{h_1/c}(z_1) \, 
\left[ D_{h_2/d} (z_2) \delta (\Delta \varphi - \pi)  \right] \,  
 \frac{\phi_{a/N}({x}_a)\phi_{b/N}({x}_b)}{{x}_a{x}_b\, {S}^2 } \, 
\nonumber \\[.5ex] 
&& \times  \; 
2 \pi \alpha_s^2 |\overline {M}_{ab\rightarrow cd}|^2 \; .
\label{double}
\end{eqnarray}
In Eq.~(\ref{double}) $K=2$ is a next-to-leading order $K$-factor, 
$x_{a,b}=p_{a,b}/p_{N_a,N_b}$ are the momentum fractions of the 
incoming partons and $z_{1,2} = p_{h_1,h_2}/p_{c,d}$ are the 
momentum fractions of the hadronic fragments. We use standard 
lowest order Gluck-Reya-Vogt PDFs~\cite{Gluck:1998xa} and  
Binnewies-Kniehl-Kramer FFs~\cite{Binnewies:1994ju}. Renormalization,
factorization and fragmentation scales are suppressed everywhere 
for clarity.  The spin (polarization) and color averaged 
matrix elements $|\overline {M}_{ab\rightarrow cd}|^2$ are given
in~\cite{Owens:1986mp}.

\begin{figure}[!t]
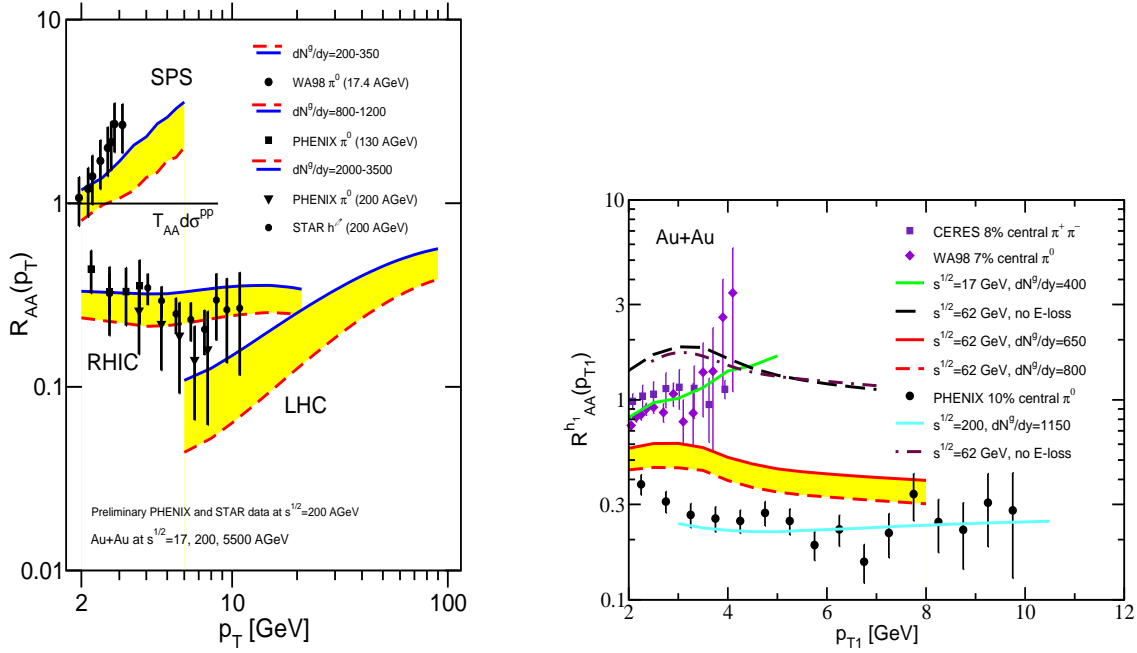

\begin{center}
\hspace*{-.75cm}
\includegraphics[width=2.4in,height=3.4in,angle=0]{Quench-200GeV.eps}
\hspace*{1cm}
\includegraphics[width=3.in,height=2.4in,angle=0]{Quench-62GeV.eps}
\caption{ Left panel: nuclear modification factor 
$R^{h_1}_{AA}(p_T)$ for single inclusive pion production at 
SPS, RHIC and the LHC. Right panel: the same nuclear modification 
for the RHIC Au+Au run at c.m. energy 
of 62~AGeV.}
\end{center}
\label{inclusives}
\end{figure}

Dynamical nuclear effects in multi-particle production can be 
studied through the ratio~\cite{Qiu:2004da}
\begin{equation}
\!  R^{(n)}_{AB} = \frac{d\sigma^{h_1 \cdots h_n}_{AB} / 
dy_1 \cdots dy_n d^2p_{T_1} \cdots d^2p_{T_n}} 
{\langle N^{\rm coll}_{AB} \rangle\, d\sigma^{h_1 \cdots h_n}_{NN} / 
dy_1 \cdots dy_n d^2p_{T_1} \cdots d^2p_{T_n}} \; .
\label{multi}
\end{equation}
Centrality dependence is implicit in Eq.~(\ref{multi}) and the 
modified cross section per average collision 
$d\sigma_{AB}^{h_1 \cdots h_n}/\langle N^{\rm coll}_{AB} \rangle $ 
can be calculated from  Eqs.~(\ref{single}) and (\ref{double})
including the numerically evaluated energy loss. 
Calculation of the nuclear modification in central Au+Au 
collisions~\cite{Vitev:2002pf} 
at $\sqrt{s_{NN}}=17, 200, 5500$~GeV are shown in the 
left panel of Fig.~(\ref{inclusives}). The right panel of
Fig.~(\ref{inclusives}) extends the GLV jet quenching 
predictions~\cite{Vitev:2004gn} to the intermediate 
RHIC energy of  $\sqrt{s_{NN}}= 62$~GeV. In this case 
parton energy loss, driven by the medium density $dN^g/dy$, 
is smaller but its effect at high $p_T$ is amplified by the 
steepness of the underlying partonic spectra.  
Similar results for the light pion suppression were 
found in~\cite{Adil}.  Baryons, however, are expected 
to be significantly less
suppressed. In fact, the $p/\pi^+$ ratio was predicted to 
be larger than at the top RHIC energy~\cite{Greco:2004yc}. 
Relating the observed high $p_T$ hadron 
attenuation~\cite{Levai:2001dc,Adcox:2001jp,Adams:2003im,Xu:2004ea}   
to the  energy density in central Au+Au collisions 
through the assumption of local thermal 
equilibrium we find that at times $ \tau_0 \simeq 0.6$~fm 
$\epsilon_0 \simeq 15 - 20$~GeV/fm$^3$ at $\sqrt{s_{NN}}=200$~GeV,
well in excess of the critical energy for deconfinement 
($\epsilon_c \simeq 1$~GeV/fm$^3$).

More detailed information about the properties of the quark-gluon 
plasma can be extracted via two particle correlation analysis.
At present,  a key question for perturbative QCD phenomenology 
is whether the medium induced gluon bremsstrahlung can significantly 
alter the di-hadron correlations measured at 
RHIC~\cite{Adler:2002tq,Rak:2004gk}. We naturally 
focus on  the away-side $| \Delta \varphi | \geq \frac{\pi}{2}$ 
case, where medium effects are the  largest.  
A modification that does not change the $\Delta \varphi$-integrated 
cross section is vacuum- and medium-induced  acoplanarity. 
The deviation of jets from being  back-to-back in a plane 
perpendicular to the collision axis arises from the soft gluon 
radiation and transverse momentum diffusion in dense nuclear 
matter~\cite{Qiu:2003pm}. In the approximation of collinear 
fragmentation, the width of the away-side hadron-hadron
correlation function  can be related to the accumulated di-jet 
transverse momentum squared in the $\varphi$-plane,   
$ \sin \sqrt{\frac{2}{\pi}} \sigma_{\rm Far} = 
\sqrt{ \frac{2}{\pi}  \langle k_T^2 \rangle_\varphi} / p_{\perp_d}$. 
Assuming a Gaussian form, 
\begin{equation}
f_{\rm vac. \; or \; med.} ( \Delta \varphi ) = 
\frac{1}{ \sqrt{2\pi} \sigma_{\rm Far} } 
e^{-\frac{(\Delta \varphi -\pi)^2}{2\sigma^2_{\rm Far}} } \;,
\label{gauss}
\end{equation}
a good description of $|\Delta \varphi| \geq \frac{\pi}{2}$ 
correlations measured in elementary p+p collisions at 
$\sqrt{S} = 200$~GeV~\cite{Adams:2003im} requires a large 
$\langle k_{T\; {\rm vac}}^2 \rangle_\varphi =  5\; {\rm GeV}^2$  
for the di-jet pair with  away-side scattered quark (and a 2.25 
larger value for a scattered gluon). Additional broadening arises 
from the interactions of the jet in the QGP that ultimately
lead to the reported energy loss,  
$ \langle k_{T\,\rm hot}^2 \rangle = 
2 {\mu^2_D(z_0)}/{\lambda_{q,g}(z_0)} \ln \frac{L}{z_0}$,
although only half  is projected on the $\varphi$-plane, 
$\langle k_T^2 \rangle_\varphi = 
\langle k_{T \; {\rm vac}}^2 \rangle_\varphi  
+ \frac{1}{2} \langle k_{T\; \rm hot}^2 \rangle$.

\begin{figure}[t!]
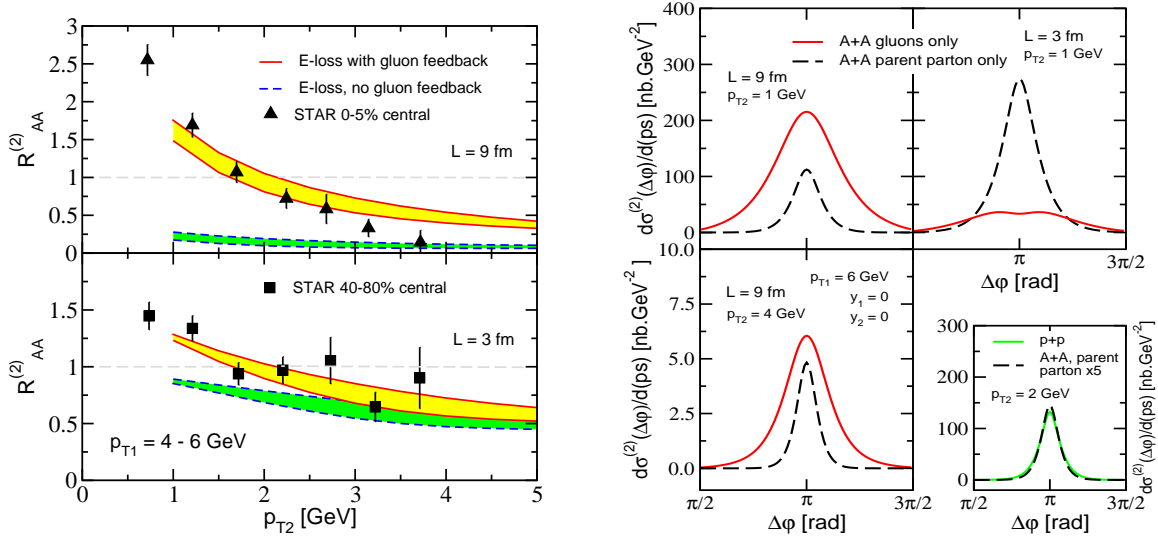

\includegraphics[width=2.8in,angle=0]{fig2.eps}
\hspace*{0.8cm}
\includegraphics[width=2.8in,height=2.8in,angle=0]{fig3.eps}
\caption{Left panel:  nuclear modification 
$R^{(2)}_{AA}(p_{T_1},p_{T_2})$ in central and 
peripheral Au+Au collisions at RHIC of the back-to-back
di-hadron correlations with and without the contribution of medium-induced 
bremsstrahlung. Right panel: angular correlations of hadrons 
dominated by large angle gluon emission. 
Solid and dashed lines give {\em separately} 
the contribution of the radiative gluons and the 
attenuated parent  parton. }
\label{yields}
\end{figure}

Two competing mechanisms do, however, change the $p_{T_2}$ dependence 
of the perturbative cross section, Eq.~(\ref{double}). First is the 
the parent jet ``d'' fractional energy loss 
$\epsilon = \Delta E_d / E_d$,  which we here for 
simplicity consider on average and evaluate by integrating 
Eq.~(\ref{unintspect}). It leads to a rescaling of 
the hadronic fragmentation momentum fraction  
$z_2 \rightarrow z_2 / (1 - \epsilon )$~\cite{Gyulassy:2003mc}.
If the energy loss is large, a second 
mechanism is invoked as a consequence. Hadronic fragments of the 
radiative gluons will increase the probability of finding  
low and  moderate  $p_{T_2}$ particles associated with the 
interacting jet~\cite{Pal:2003zf}.

To calculate di-hadron correlations, we first map the jet structure 
of a hard $90^0$-\,scattered parton on  rapidity  
$y \approx \eta = - \ln \tan (\theta /2)$  ($\theta$ being the angle 
relative to the  beam axis) and azimuth $\phi$,   
$$ \tan^2 \theta^* = \cot^2  \theta + \tan^2 \phi \, , \; 
\tan \delta  = - \frac{ \cot \theta}{\tan \phi }  \;, \;  
    \left| \frac{\partial(\sin \theta^*, \delta ) }
    {\partial(\theta , \phi )}  \right|
    = \frac{1}{\sin^2 \theta  \cos^2 \phi}
    \frac{ (\tan^2 \phi + \cot^2  \theta )^{-1/2} } 
  {( 1 + \tan^2 \phi + \cot^2 \theta )^{3/2} }     \;. $$
The approximately flat rapidity distribution of the away-side jet 
near $y_2=0$ can be used to sum over all emission angles 
$\theta \in (\theta_{\rm min}, \theta_{\rm max}) \subset 
(0,\pi)$ yielding 
\begin{equation}
\frac{dN^g_{\rm med}}{d \omega d \phi} = 
\int\limits_{\theta_{\min}}^{\theta_{\max}}  d \theta  \;
\left[ \, \frac{dN^g_{\rm med}}{d \omega d \sin \theta^* d \delta }  
\left| \frac{\partial(\sin \theta^*, \delta ) }{\partial(\theta , \phi )} 
  \right| \; \right]  \; .
\label{trans}
\end{equation} 
It is critical to note that projection on a plane 
coincident with the jet cone axis  (the $\phi$-plane in 
Eq.~(\ref{trans}) is one example) efficiently masks the  
$\theta^* \rightarrow 0$  void in the angular distribution of 
medium-induced gluons reported in Fig.~\ref{hole}. Our conclusion 
is independent of the physical mechanism that depletes the 
parton (or particle) multiplicity in a cone around the jet axis.

The end analytic result for the modification 
to Eq.~(\ref{double}) per average nucleon-nucleon collision 
in the heavy ion environment  can be derived  from the energy 
sum rule for all hadronic fragments from the jet,
\begin{eqnarray}
D_{h_2/d} (z_2) \delta (\Delta \varphi - \pi)  \; \; & \Rightarrow &  
 \frac{1}{1-\epsilon} D_{h_2/d}\left( \frac{z_2}{1-\epsilon} \right) 
f_{\rm med.}( \Delta \varphi )  
+  \, \frac{p_{T_1}}{z_1}    \int_0^1 \frac{d z_g}{z_g}  
D_{h_2/g}(z_g) 
\nonumber \\  && 
   \times \int_{-\pi/2}^{\pi/2} d \phi  \;
\frac{ dN^g_{\rm med} (\phi) }{ d\omega  d  \phi }  
f_{\rm vac.}( \Delta \varphi - \phi) \;. 
\label{nucmod}
\end{eqnarray}
Here, $z_g = p_{T_2} / \omega $.

Numerical results,  shown in the left panel of Fig.~\ref{yields}, 
correspond  to triggering on  a high $p_{T_1} =\;$4 - 6~GeV 
pion and measuring all associated $\pi^+ + \pi^0 + \pi^-$.  
Depletion of hadrons from  the quenched parent parton 
alone leads to a large suppression of the  double inclusive 
cross section -- a factor of 5 - 10 in central
and a factor of 1.5 - 2 in peripheral reactions with weak 
$p_{T_2}$ dependence. Hadronic feedback from the  
medium-induced gluon radiation, however, completely changes 
the nuclear modification factor $R^{(2)}_{AA}$.     
It now shows a clear transition from a quenching of 
the away-side jet at high transverse momenta to enhancement
at  $p_{T_2} \leq 2$~GeV, a scale significantly 
larger than the one found in~\cite{Pal:2003zf}.  
STAR data in central and peripheral Au+Au 
collisions~\cite{Adler:2002tq} is shown for comparison.

Two-particle distributions in A+A reactions,  
calculated  from Eqs.~(\ref{double}) and (\ref{nucmod}) 
for a $p_{T_1} = 6$~GeV trigger pion and two different 
$p_{T_2}=1,\;4$~GeV associated pions,  are shown in 
the Fig.~(\ref{yields}).  Qualitatively, the medium-induced 
gluon component to the cross section controls the growth 
of the correlation width in  central and semi-central nuclear 
collisions. Quantitatively, the effect should be even larger than    
the one estimated here, which is limited by the 
imposed  $ 0 < \theta^* < \frac{\pi}{2} $ constraint. 
Experimental measurements of significantly 
enhanced widths for $| \Delta \varphi| \geq \frac{\pi}{2}$  
two-particle correlation in A+A collisions should thus point 
to copious hadron production from medium-induced  large angle 
gluon emission.

\section{Dynamical shadowing and energy loss in cold nuclear matter} 

Single and double inclusive hadron production in reactions involving
nuclei is also modified by dynamical shadowing effects. Suppression of
the particle production rates arises from the coherent final state 
scattering of the struck small-x parton with several nucleons and the
generation of dynamical parton mass $m^2_{dyn} = \xi^2 A^{1/3}$. Here, 
$\xi^2 = 0.09 - 0.12$~GeV$^2$ for quarks is the scale of higher 
twist extracted from deeply inelastic scattering in heavy 
targets~\cite{Qiu:2003vd}. For electroweak vector meson exchange 
processes in DIS resumming the power corrections leads to an 
effective rescaling in the values of Bjorken-x.  For p+A and A+A 
reactions the dynamical mass is amplified for the case of final 
state gluon scattering  by the Casimir ratio $C_A/C_F = 2.25$.
It leads to significantly larger power 
correction effects in hadronic reactions. Effectively,
\begin{equation} 
x_B \rightarrow x_B \left( 1 + \frac{\xi^2(A^{1/3}-1)}{Q^2}  \right) 
\; , \; \; \quad 
x \rightarrow x \left( 1 + \frac{C_d\xi^2(A^{1/3}-1)}{-\hat{t}} \right) \; ,
\label{shift}
\end{equation} 
where the normalization relative to the proton is already included 
in Eq.~(\ref{shift}) and  $C_d = 1 \;(2.25)$ for quarks (gluons) 
respectively.

\begin{figure}[t!]
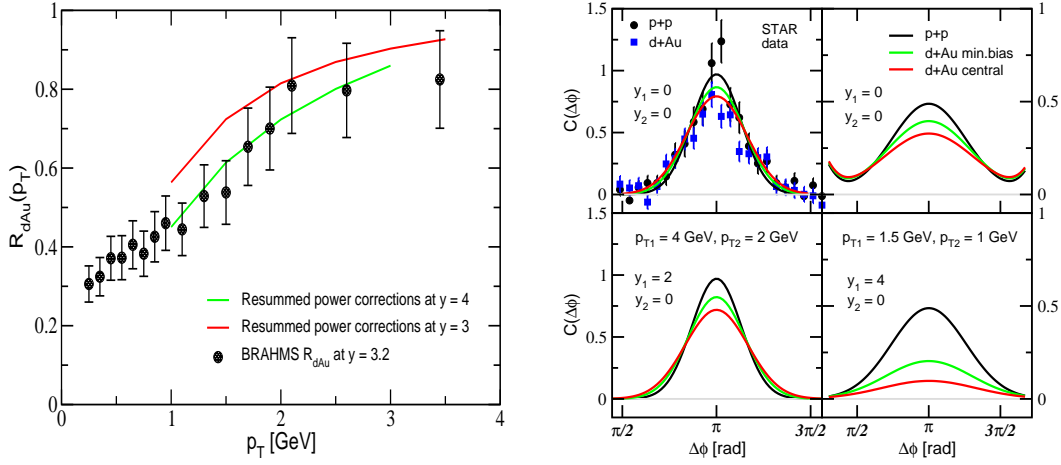

\begin{center}
\includegraphics[width=2.6in,height=2.4in,angle=0]{Comparison.eps}
\hspace*{.5cm}
\includegraphics[width=2.6in,height=2.4in,angle=0]{Fig3.eps}
\end{center}
\caption{Left panel: modification of the single inclusive hadron
production from resummed nuclear enhanced power corrections in 
forward rapidity d+Au collisions at RHIC. Right panel: modification 
of the back-to-back di-hadron correlations versus rapidity, centrality
and $p_T$ in $\sqrt{s_{NN}}=200$~GeV d+Au reactions.}
\label{twist}
\end{figure}

The left panel of Fig.~\ref{twist} shows the nuclear modification
$R_{dA}(p_T)$ for hadron production in d+Au reactions~\cite{Qiu:2004da}. 
We emphasize that the effect is large at small $p_T$ 
and relatively small at high $p_T$. Note that at forward rapidity 
in Eq.~(\ref{shift})  $-\hat{t} = p_T^2$. 
Data on forward $y$ hadron production from BRAHMS is also 
shown~\cite{Arsene:2004ux}.  The right panel of Fig.~\ref{twist}
presents the broadening and attenuation of the two particle 
back-to-back correlations.  Such effects are found to be large at 
forward  rapidity and at small $ p_{T_1}, p_{T_2} $~\cite{Qiu:2004da}.  
The coherent multiple scattering  discussed here take place (in the 
c.m. of the collision) over a short interval $ t \sim R_A/\gamma  \ll R_A $  
and for the case of A+A reactions  precedes and complements 
the energy loss in the QGP which develops at time scales  
$ t \sim R_A $.

In cold nuclear matter energy loss is anticipated to be 
significantly smaller when compared to the one in the quark-gluon
plasma. It has been experimentally verified through measurements 
in $d+Au$ reactions at RHIC at $\sqrt{s_{NN}}=200$~GeV where a 
small enhancement consistent with Cronin multiple 
scattering~\cite{Vitev:2003xu} has been observed~\cite{Arsene:2003yk}. 
It as been argued, however, that the radiative 
energy loss induced by the scattering of fast on-shell partons 
in nuclear matter evaluated by Bertsch and Gunion~\cite{Gunion:1981qs} 
can be implemented as a ``Sudakov'' suppression  factor 
$ S(x_F) \approx 1 - x_F $~\cite{Kopeliovich:2005ym} and can give
a good description of the $A^\alpha,\; \alpha <1 $ suppression
in a large class of observed cross sections in hadronic 
reactions at forward~$x_F$.

\section{Conclusions}

In summary, we calculated the nuclear modification of inclusive
moderate and high $p_T$ particle production and the back-to-back particle
correlations  in the  framework of the perturbative QCD 
factorization approach,  augmented by inelastic jet 
interactions in the quark-gluon plasma.  Jet tomographic  
analysis of hadron attenuation points to initial energy
density of $15 - 20$~GeV/fm$^3$ in $\sqrt{s_{NN}}=200$~GeV
central Au+Au collisions. 
At RHIC energies we found that the  medium-induced gluon radiation  
determines the $| \Delta \phi | \geq \frac{\pi}{2}$  
two-particle yields and the width of 
their correlation function to surprisingly high transverse 
momentum  $p_{T_2} \sim 10$~GeV.  The predicted transition
from back-to-back jet enhancement to  back-to-back jet quenching 
is quantitatively compatible with the recent STAR data.
In cold nuclear matter at RHIC coherent final state 
multiple scattering gives the dominant contribution 
to the attenuation of single and double inclusive hadron 
production. Energy loss, however, may also play an important 
role at forward~$x_F$.

\vspace*{+.2cm}
   
\noindent {\bf Acknowledgments:} 
This work is supported by the 
J.~R.~Oppenheimer Fellowship  of the Los Alamos National 
Laboratory and by the US Department of Energy.

\end{document}